# Enhanced radiation damage tolerance of amorphous interphase and grain boundary complexions in Cu-Ta

Doruk Aksoy [1], Penghui Cao [1,2], Jason R. Trelewicz [3], Janelle P. Wharry [4], Timothy J. Rupert [1,2,*]

[1] Department of Materials Science and Engineering, University of California, Irvine, CA 92697, USA
[2] Department of Mechanical and Aerospace Engineering, University of California, Irvine, CA 92697, USA
[3] Department of Materials Science and Chemical Engineering, Stony Brook University, Stony Brook, NY 11794, USA
[4] School of Materials Engineering, Purdue University, West Lafayette, IN 49707, USA
[*] Corresponding author: trupert@uci.edu

**ABSTRACT**

Amorphous interfacial complexions are particularly resistant to radiation damage and have been primarily studied in alloys with good glass-forming ability, yet recent reports suggest that these features can form even in immiscible alloys such as Cu-Ta under irradiation. In this study, the mechanisms of damage production and annihilation due to primary knock-on atom collisions are investigated for amorphous interphase and grain boundaries in a Cu-Ta alloy using atomistic simulations. Amorphous complexions, in particular amorphous interphase complexions that separate Cu and Ta grains, result in less residual defect damage than their ordered counterparts. Stemming from the nanophase chemical separation in this alloy, the amorphous complexions exhibit a highly heterogeneous distribution of atomic excess volume, as compared to a good glass former like Cu-Zr. Complexion thickness, a tunable structural descriptor, plays a vital role in damage resistance. Thicker interfacial films are more damage-tolerant because they alter the defect production rate due to differences in intrinsic displacement threshold energies during the collision cascade. Overall, the findings of this work highlight the importance of interfacial engineering in enhancing the properties of materials operating in radiation-prone environments and the promise of amorphous complexions as particularly radiation damage-tolerant microstructural features.

# 1. INTRODUCTION

Next-generation damage-resistant materials for nuclear applications must be designed to withstand ultra-high temperatures and radiation doses [1], while also featuring improved structural and functional properties [2,3]. Despite continued efforts, the evolution of microstructure during service often leads to material damage such as irradiation-induced swelling and embrittlement, which continue to limit the lifetime of engineering materials operating in these environments. Irradiation causes high-energy particles to transfer kinetic energy to primary knock-on atoms (PKAs) [4], leading to the formation of collision cascades characterized by profuse displacements of atoms from their parent lattice positions. This chain of events typically drives the formation of various extended defects such as point defect clusters [5], dislocation loops [6], and stacking fault tetrahedra [7]. These defects act as precursors to deleterious behavior, including embrittlement [8]. The formation of radiation-induced defect structures can lead to material failure when subjected to mechanical stress, and additional events such as the creation of transmutation products (e.g., He atoms leading to bubble formation) can also occur [9].

Interface-dominated materials, such as nanostructured metals and nanoscale layered films, display enhanced radiation tolerance due to a significant density of grain boundaries that serve as defect sinks capable of mitigating radiation-induced defects, as evidenced by the formation of denuded zones near the boundary [10]. A critical topic is the change in interfacial structure due to defect and solute segregation [11,12], which in turn can significantly impact sink efficiency. Influenced by various factors, including crystallographic orientations and interfacial segregation spectra, defect and solute segregation can modify grain boundary structures, ultimately altering their ability to absorb and mitigate defects [13–19]. Grain boundaries in polycrystalline materials range from ordered high-symmetry to high-energy disordered structures [20], providing a diverse set of potential sites for defect absorption that can minimize radiation damage [21]. However, grain boundaries can also become saturated under prolonged radiation exposure, impairing their healing capacity [22]. Regions with excess atomic volume, such as those from misfit dislocations in face-centered cubic (FCC)/body-centered cubic (BCC) phase boundaries [23,24] and amorphous complexions [25], offer improved radiation damage healing through enhanced defect recombination and annihilation. This healing can be further enhanced by tailoring the degree of radiation-induced mixing, or the redistribution of segregated solutes between the amorphous complexion region and the bulk region [26].

Immiscible nanostructured Cu-Ta alloys can act as a unique class of interface-dominated materials [27], serving as an important proving ground for the study of radiation damage mitigation. Recent work has shown that both grain boundaries (i.e., separating two Cu crystals) and interphase boundaries (i.e., separating a Cu crystal from a Ta crystal) provide a plethora of potential defect sink sites that facilitate recombination of irradiation defects [28–32]. Amorphous interphase regions can also form during some irradiation conditions, such as $Xe^+$ ion irradiation at room temperature, which yields a ~50 nm amorphous interphase layer in Cu-10 at.% Ta after a dose of 10 displacements per atom (dpa), as shown in Fig. 1. The insets in Fig. 1 show the diffraction patterns, which indicate an increase of amorphous phase at the 10 dpa dose level. The amorphous complexions first formed at lower radiation doses and thickened to the state shown here. A more detailed experimental report, to be published elsewhere, will describe irradiation with various ion species, at different temperatures, and with increasing dpa levels. As these interfacial structures are created by irradiation, they serve as a unique example of microstructural features formed during service conditions that can mitigate further damage accumulation. Cu-Ta alloys have already been shown to exhibit elevated thermal stability and mechanical strength [27,33–35], so this unique capacity to host disordered interfaces alongside other types of grain boundary and interphase structures holds great promise for radiation damage resistance. Different from typical metallic glasses, glasses in the Cu-Ta system have a nanophase separated microstructure at room temperature, with Ta nanoparticles suspended in an amorphous Cu matrix that resist permanent aggregation [36,37]. Overall, the Cu-Ta alloy's unique structural attributes represent an opportunity for interface-dominated materials capable of combating radiation damage.

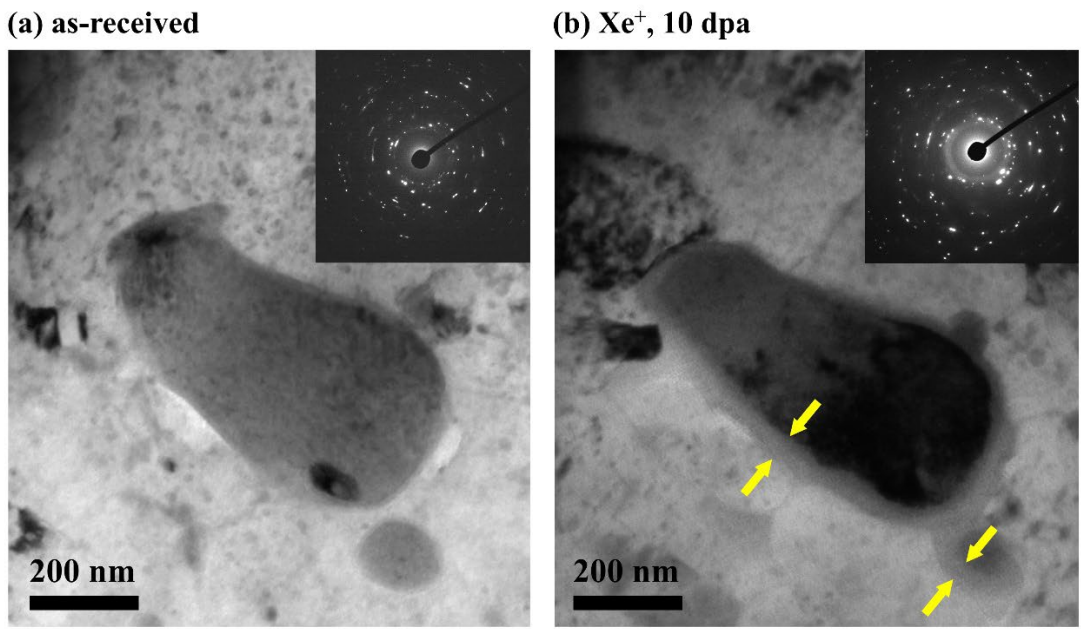

**Fig. 1. Transmission electron microscopy (TEM) images of Cu-10 at.% Ta samples in the (a) as-received condition and (b) after 10 dpa $Xe^+$ ion irradiation at room temperature, with yellow arrows indicating amorphous interphase layers.** The insets in Figure 1 show the diffraction patterns, which indicate an increase of amorphous phases at the 10 dpa dose level.

In this study, the defect annihilation behavior and damage tolerance of amorphous interfacial complexions within Cu-Ta are investigated via PKA collision cascade simulations. Four types of model interfaces are used for this comparison, including an ordered grain boundary, an ordered interphase boundary, and amorphous complexions at Cu-Cu and Cu-Ta interfaces. This work represents the first study to distinguish between amorphous complexions in grain boundaries and interphase boundaries in terms of their ability to dissipate radiation damage. Our findings show that amorphous complexions exhibit minimal residual defect content compared to their ordered counterparts, highlighting the importance of interfacial disorder for radiation tolerant microstructures. The amorphous interphase complexion connecting Cu and Ta is found to result in the least residual defect damage, in addition to being an unbiased sink, with the amorphous complexion separating Cu grains also exhibiting excellent performance. The simulations reveal that the radiation damage healing behavior is significantly influenced by the availability of free volume in the amorphous complexions. The impact of film thickness on radiation damage generation and mitigation behaviors is explored as well, with thicker films shown to provide an improved barrier to the collision cascade that reduces the amount of damage generated by changing the defect production rate relative to sharper interfaces. The disordered regions in these models contribute to collision cascade confinement and provide more opportunities for recombination due to their higher excess atomic volume. As a whole, the findings of this study show that disordered interfacial states should be integrated into microstructures for increased resistance to radiation damage.

## 2. METHODS

Four interface types were modeled in this study: (i) ordered grain boundary, (ii) ordered interphase boundary, (iii) amorphous grain boundary (AGB), and (iv) amorphous interphase boundary (AIB). In addition, a single crystal model was simulated to provide a baseline reference.

All five models are shown in Fig. 2, with yellow atoms representing Cu and green atoms representing Ta. The Large-scale Atomic/Molecular Massively Parallel Simulator (LAMMPS) software [38] was employed for molecular dynamics (MD) simulations with an angular-dependent Cu-Ta interatomic potential [39] that accurately reproduces a wide range of properties, especially in configurations with significant lattice distortion that are critical for simulating radiation damage. Notably, this potential has been optimized to model point defect properties such as formation energies and diffusion barriers. For instance, the predicted vacancy migration energy matches first-principles calculations, although both slightly overestimate experimental values. Similarly, the formation energy for certain interstitial orientations is in close agreement with first-principles data. These assessments indicate that the potential is reliable for simulating the behavior of point defects. This potential also predicts the nanophase separated structure associated with Ta-doped grain boundaries and Cu-Ta glasses at lower temperatures [37].

Previous studies suggest that the initial heat spike resulting from energy transfer during early ballistic collisions dissipates in a few picoseconds [40,41]. To control this sudden spike and manage the collisions of high-energy atoms, the maximum distance an atom could move was limited to 0.05 Å per timestep using an adaptive timestep, as regularly performed in the literature (see, e.g., [7,21]). This adaptive timestep approach works synergistically with the well-parameterized nature of the selected Cu-Ta interatomic potential, which employs a cubic spline-based angular-dependent potential (ADP) formulation that allows for precise modeling of potential energy as a function of distance. This formulation is particularly effective in accurately simulating short-range interactions without requiring additional Ziegler–Biersack–Littmark terms, as discussed in Ref. [41].

All simulation models used periodic boundary conditions in all directions. The single crystal model was oriented in the same direction as the upper lattice of the grain boundary model, since the collision cascade will extend into the upper grain, and contained ~160,000 atoms with approximate dimensions of 11.4 nm × 22.8 nm × 7.2 nm. The grain boundary model consisted of two high-angle grain boundaries arranged in a periodic array, with both being Σ5 [100] (031) symmetric-tilt interfaces. The equilibrium grain boundary structure was obtained using the methodology described in Ref. [20] and had the approximate dimensions of 11.5 nm × 23.0 nm × 7.3 nm, containing 159,200 atoms. The well-established kite structure [42] was observed for the grain boundary model, as shown in the inset of Fig. 2(b).

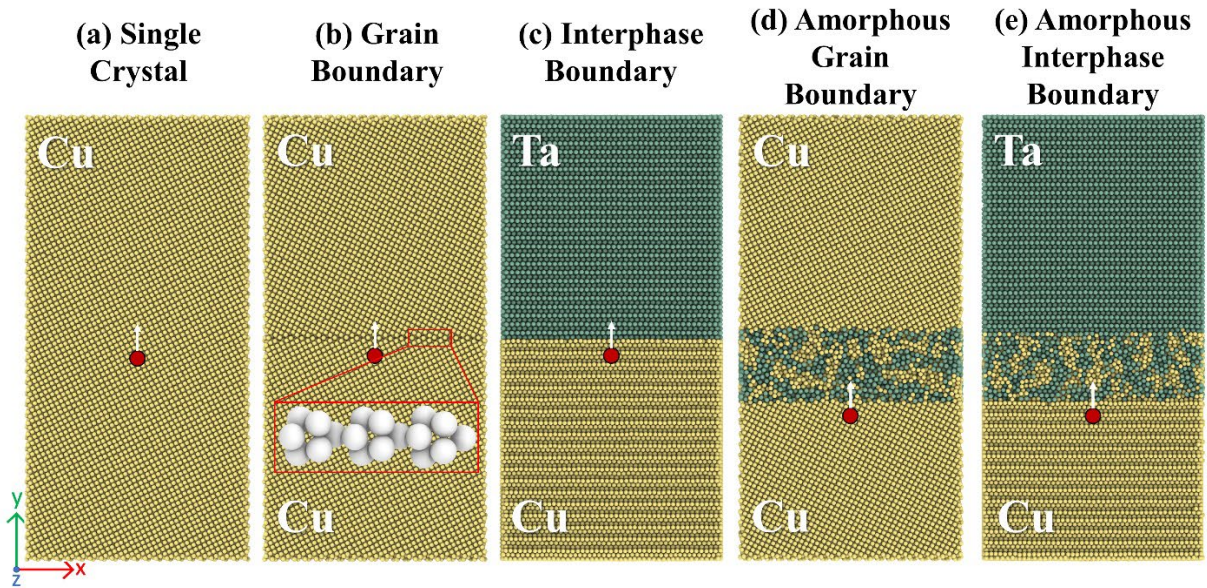

**Fig. 2. The simulation models include (a) single crystal, (b) ordered grain boundary, (c) interphase boundary, (d) amorphous grain boundary, and (e) amorphous interphase boundary. The kite structure associated with the Σ5 [100] (031) symmetric-tilt grain boundary is shown in detail in the inset of (b). The PKA location is shown as a red dot in each image 0.5 nm away from the interface (except (a) for the single crystal), where the atom is moving upward along the Y-axis. Cu atoms appear yellow and Ta atoms appear green in all configurations.**

The coincident site lattice model is inadequate for obtaining stable structures in interphase boundaries due to phase structure mismatch [43]. Instead, a quasiperiodic pattern based on the Kurdjumov-Sachs orientation relation (i.e., ⟨110⟩ {111}$_{Cu}$//⟨111⟩ {110}$_{Ta}$) dictates the structure formed by stacking planes of close-packed FCC and BCC phases [44]. This choice was motivated by the low lattice misfit between the two phases, resulting in reduced strain energy in the boundary compared to other common orientation relations such as the Nishiyama-Wasserman and Greninger-Troiano relations [45]. Fig. 3(a) shows the starting configuration of the interphase boundary from a viewing direction along the boundary normal, with one atomic layer shown from either side of the two parallel planes (i.e., one (111) plane from FCC Cu and one (110) plane from BCC Ta). The black atoms correspond to those residing in the terminal Cu plane that are near-coincident with the atoms residing in the terminal Ta plane, which forms a quasiperiodic pattern. Introducing vacancies at the intersections of this pattern followed by relaxation of the system has been shown to reduce the total energy of the interphase boundary [23], so such a procedure is followed here. Fig. 3(b) displays the final structure after relaxation, with a magnified view presented to the right. The final simulation cell had approximate dimensions of 11.8 nm × 24.0 nm × 7.3 nm and contained ~144,500 atoms.

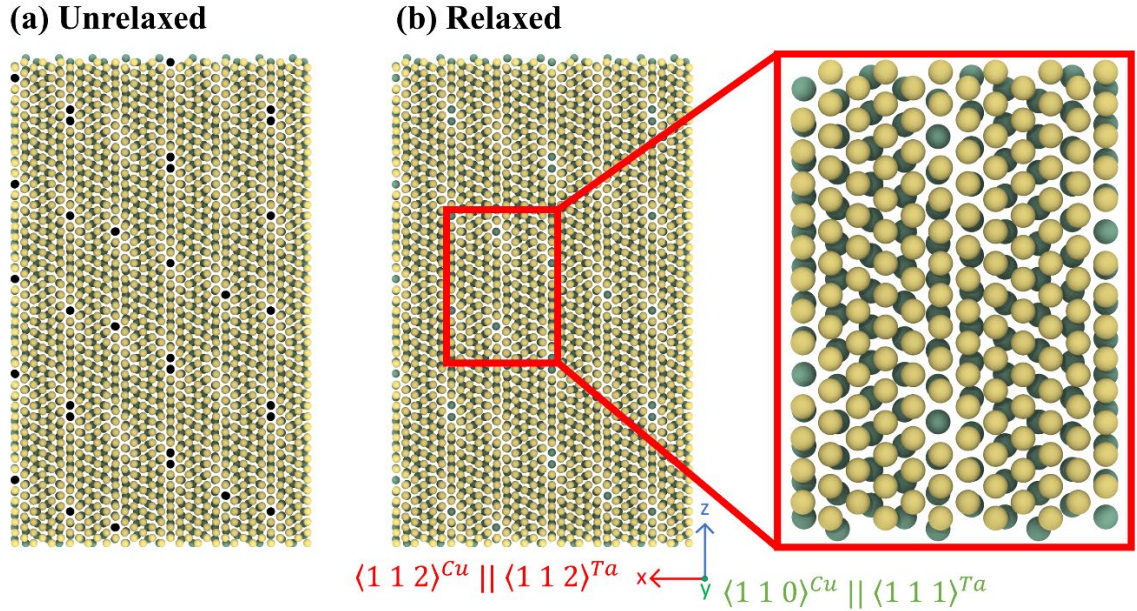

**Fig. 3. (a) The Kurdjumov-Sachs orientation relation was used for creating the interphase boundary, where one atomic layer from each phase is shown and viewed along the grain boundary normal direction. Black atoms correspond to the near-coincident atoms between the two adjacent planes. (b) The minimum energy configuration is obtained by introducing vacancies at near-coincident sites, followed by relaxation. The callout provides a magnified view of atomic positions in the region outlined in red.**

The amorphous complexion models (AGB and AIB) were subsequently created from the associated ordered grain boundary and interphase boundary structures. For the AGB model, the amorphous film was created in a region centered on the original grain boundary. In this case, Ta atoms were intentionally incorporated into the Cu-based grain boundary to recreate the phenomenon of segregation-induced grain boundary pre-melting [46], which is enabled by the chemically heterogeneous atomic environment. Hsieh and Balluffi [47] demonstrated that grain boundaries in pure metals remain solid even at 99.9% the melting point. In contrast, Luo and Shi [48,49] have shown that grain boundary pre-melting can occur at temperatures of roughly 60-85% of the melting temperature for doped grain boundaries. For the AIB model, a slab at the interphase boundary was extended into the Cu lattice, because the initial knock-on atom was pointed toward the Ta lattice and this choice ensured that there was sufficient room in the Ta crystal for the collision cascade to occur. To maintain compositional consistency between the two types of amorphous complexions, half of the Cu atoms in the slab regions were randomly replaced with Ta atoms. This doping also ensured that the amorphous structures were within the metastable liquid miscibility gap regime for the Cu-Ta system. In this composition range, the Cu atoms within the slab melt first when there is a transition from crystalline to non-crystalline form, while the Ta atoms are suspended in liquid Cu [50,51]. To accomplish this, the slab was heated to 1600 K in the span of 200 ps using an isobaric-isothermal ensemble (NPT), while fixing the atoms outside of the slab. Subsequently, the slab was quenched in two steps, first from 1600 K to 650 K and then from 650 K to 300 K, each over 200 ps using an NPT ensemble. Now solid, all atoms in the system were subjected to a thermal equilibration procedure at 300 K for 200 ps, allowing the overall system to stabilize into two crystals separated by a nanophase separated amorphous structure. The initial amorphous complexion thicknesses were chosen to be 3 nm, and subsequent simulations with varying thickness between 2 and 4 nm were carried out to test the effect of complexion thickness on damage tolerance. Figs. 1(d) and 1(e) show the final relaxed atomic configurations for the 3 nm thick AGB and AIB models, respectively. The AGB model had approximate dimensions of 11.6 nm × 23.3 nm × 7.4 nm and contained 159,200 atoms, while the AIB model had approximate dimensions of 12.0 nm × 24.3 nm × 7.4 nm and contained ~144,500 atoms.

After all simulation cells were created, a conjugate gradient energy minimization step was applied to each model, followed by an MD simulation under an NPT ensemble to ensure zero hydrostatic pressure at 300 K for a total of 30 ps. Next, the PKA atoms were selected 0.5 nm away from the interface (perpendicular to grain boundary, interphase boundary, or amorphous-crystalline interface plane). Ten PKA atom locations were chosen to perform different, yet thermodynamically-equivalent collision cascade simulations. These PKA atoms shared the same X and Y locations, but different positions along the Z-axis (i.e., through the model thickness). The selected simulation cell dimensions were at least 50% larger than the size of the maximum damage in all directions to prevent the collision cascade from interacting with its own periodic images. PKA events were initiated by applying a prescribed kinetic energy to the selected atom (red atoms in Fig. 2) in the direction of the interface (black arrows). A PKA energy of 2 keV was chosen for the majority of this study, which falls within the range where PKA energy is linearly related to the creation of Frenkel pairs [52]. The effect of PKA energy on damage in the AGB model was also investigated with additional simulations at higher PKA energies of 5 keV and 10 keV. The edge atoms of the simulation box were maintained at 300 K during the simulation to act as a thermal sink, and the simulations were performed for 52 ps under a microcanonical ensemble after releasing the PKA. This duration was chosen based on the observation that the simulation cell temperature and the defect configurations ceased evolving by this time, ensuring that the defect evolution trend converged to a metastable state. Such an approach aligns with the understanding that only recombination during thermalization is captured and real diffusive processes over longer time scales are not accounted for, potentially leading to an overestimate of the defect production term in any rate theory equation.

The Wigner-Seitz defect analysis method [53] was used to identify defects in the crystalline regions during the simulations. This method compares against the initial unperturbed structure after creating Voronoi cells around each atom. Cells with more than one atom are interstitials, while cells without any atoms are vacancies. The primary challenge with this method when applied to interfaces is the inherent disorder and non-crystalline nature of these regions. Given the lack of a well-defined repeating lattice in such areas, the Wigner-Seitz method can often misidentify regular atomic positions as defects or vice versa. Specifically, the reconfiguration of interfaces during cascade simulations can lead to false interstitial and vacancy counts, as defining true point defect sites in a disordered interface is challenging unless one can identify regularly repeating sites. While thermal fluctuations do impact all atoms, the inherent structural disorder at the interfaces makes them particularly susceptible to misidentification in this type of analysis. To mitigate these issues and ensure a more accurate collision cascade defect count, interfacial defect atoms were excluded from the defect analysis for all models, whether they are amorphous and ordered. This exclusion was facilitated by the adaptive common neighbor analysis method included in OVITO [53], which was used to distinguish the crystalline regions from the interfaces. Additionally, the edges of the simulation cell with periodic boundary conditions were also not considered in the Wigner-Seitz defect analysis to avoid potential inaccuracies.

## 3. RESULTS AND DISCUSSION

Annealing temperature and quenching rate play a pivotal role in determining the structure and properties of glassy phases. The unique nanophase separated structure of Cu-Ta glass, influenced by these parameters, distinguishes it from alloys with higher glass-forming ability, such as the Cu-Zr system, which is among the most studied binary alloys that can be vitrified [54,55]. In situ transmission electron microscopy (TEM) and ex situ ion irradiation studies on nanocrystalline Cu-Zr demonstrated how amorphous complexions can enhance radiation tolerance [56], by reducing the density of defects formed inside the microstructure during irradiation. To illustrate the difference between the two alloy systems, bulk metallic glass Cu-Zr and Cu-Ta models were prepared by fully melting and mixing the solute and solvents, as shown in Figs. 4(a) and 4(b). The preamble for forming the discussed nanophase separation in Cu-Ta is to be within the metastable liquid miscibility gap regime, which depends on the selection of the interatomic potential. Referring to the experimental phase diagram [50], the selected conditions closely mirror the experimental range conducive to the nanophase separated microstructure. In the simulations, an equimolar Cu-Ta ratio was selected for the amorphous complexions based on the understanding that such compositions are frequently effective in forming metallic glasses. This is due to the fact that in binary alloy systems, compositions near the eutectic points, often at or close to equiatomic ratios, tend to exhibit enhanced glass-forming ability [57]. Previous studies have shown that the Cu-Ta liquid solution rapidly transitions into a nanophase separated structure [37]. This fast transformation suggests that because there is a strong driving force, likely originating from the irradiation, the amorphous complexion should reach a metastable structural state quickly. This is corroborated by the findings of this study and prior simulations, which revealed that even when an oversized particle was introduced into the colloidal structure, it disintegrated into smaller particles within 100 ps, thus maintaining the uniform colloidal structure [37]. Rapid restructuring supports the idea that the system tends to stabilize into this structure over short timescales, even when perturbed. To provide a comparison with the Cu-Zr system [25], the interatomic potential developed by Mendelev et al [58] was utilized to simulate the glassy structure in this alloy. The liquid configurations were subjected to a controlled cooling process from 2000 K to 50 K over 50 ps, resulting in solid amorphous structures for both systems. The cooling rate implemented is more than a million times faster than the rate necessary to create bulk metallic glasses, meaning that it prohibits the formation of the equilibrium crystalline phases [57].

For the quenched models, the average atomic volume of Cu atoms in the Cu-Zr and Cu-Ta bulk amorphous metal models is 33% and 15.8% higher, respectively, than the average atomic volume associated with Cu atoms in the single crystal model. These results clearly demonstrate an increase in free volume associated with the structural disorder inherent to an amorphous phase. The nanophase separation of the Cu-Ta alloy can be seen in the top frames of Fig. 4(b) where clusters rich in either Cu or Ta atoms are observed, with this effect being more prominent in the quenched configuration. The clustering of like atoms causes the Cu-Ta glass to exhibit regions of very low atomic volume in the quenched state shown in the bottom right frame of Fig. 4(b). Elemental clustering can also be seen in the radial distribution functions presented in Figs. 4(c) and 4(d). The three curves represent different atomic pairings, with the areas under each curve specified by the value shown above the peak of a given curve. The radial distribution function shows the probability of finding another atom of the specified type within a given pair separation distance. For Cu-Zr in Fig. 4(c), the glass is a relatively homogeneous mixture of the component elements, with the highest integrated intensity being seen for the Cu-Zr atom pairing. In contrast, the highest integrated intensities are for the monatomic pairs in the Cu-Ta alloy, showing that there is an increased likelihood of finding clusters of either Cu or Ta atoms (i.e., the nanophase separated structure of the Cu-Ta). This observed tendency toward monatomic pairs in Cu-Ta is consistent with the larger miscibility gap in the Cu-Ta phase diagram compared to that of Cu-Zr [50,59].

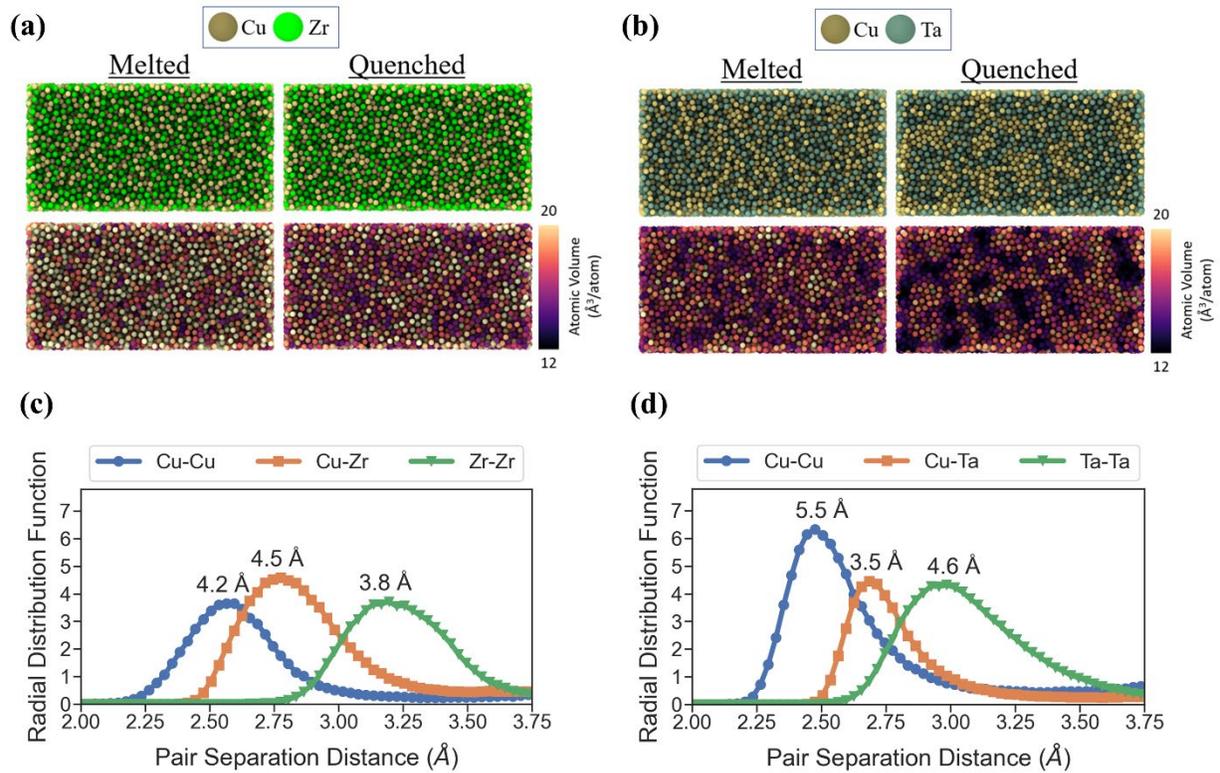

**Fig. 4.** Atomic configurations of (a) Cu-50 at.% Zr and (b) Cu-50 at.% Ta bulk amorphous models, showing the liquid (melted) states and the quenched metallic glass structures after rapid cooling. Excess atomic volumes are shown in the bottom row. Radial distribution functions measure the probability of finding a particle of a specified type within the pair separation distance for the (c) Cu-Zr and (d) Cu-Ta systems, where the integrated areas under each curve are labeled.

Armed with an understanding of the unique Cu-Ta amorphous structure, we next move to study the defect formation and accommodation process in each interfacial model. The evolution of a collision cascade can be tracked by analyzing the kinetic energy and number of defects as a function of time. Fig. 5(a) presents the kinetic energy evolution for the 3 nm thick AGB model, where only atoms with kinetic energies greater than 0.4 eV are shown. The first collection of snapshots is presented at 0.25 ps intervals (up to 1.75 ps), with simulation time shown above each image, followed by the final snapshot at the very end of the simulation (i.e., after 52 ps). The semi-transparent slab in these images represents the amorphous complexion region. The initial heat spike dissipates over approximately 1 ps, during which localized melting behavior can transpire at interfaces due to the sudden increase in local energy [60]. As the kinetic energy spreads into the system, it manifests as a local temperature rise, accompanied by the formation of Frenkel pairs. Fig. 5(b) shows the time evolution of point defect content for the 3 nm thick AGB over the same time interval. A burst of point defects is observed, reaching the highest defect content near 1 ps. A series of recombination events then follows until the system attains a local equilibrium that is invariant to further MD equilibration. The last snapshot of Fig. 5(b) displays the final defect configuration, comprised of a single residual vacancy in this case.

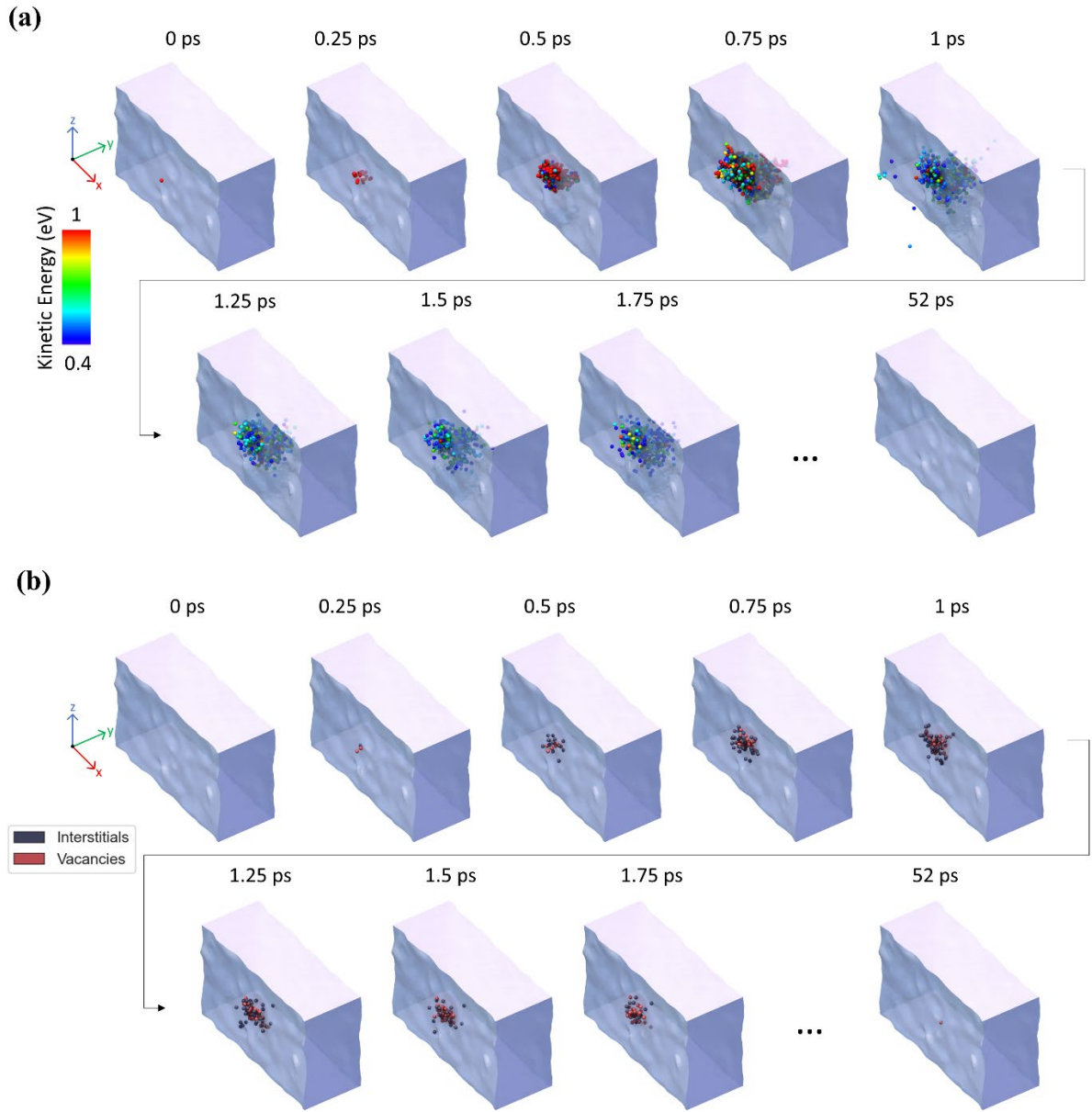

**Fig. 5.** The temporal evolution of (a) atomic kinetic energy and (b) point defect content during the collision cascade event in the 3 nm AGB model, where the amorphous complexion region is illustrated by semi-transparent gray boxes. (a) The atomic kinetic energy is shown at different time steps, with only atoms with energies greater than 0.4 eV shown. (b) The corresponding defect configurations during the cascade event are presented, with interstitials and vacancies represented by blue and red colored atoms, respectively.

The radiation damage mitigation behavior of the different interfaces is next obtained by quantifying the residual defect content at the end of each simulation. Fig. 6 shows the numbers of interstitials and vacancies corresponding to the recovered configuration for each interface. Considering variations in defect populations across the ten independent collision cascade simulations for each interface, delimiting lines represent minimum and maximum defect values, vertical box bounds indicate the 1st and 3rd quartiles, the middle lines signify median values, the points outside the distribution illustrate outlier values (the points that remain outside of 150% of the inter-quartile range [61]), and green plus sign symbols represent the mean values. Representative examples of final snapshots for each interface type are provided below the plot, where the enlarged particles represent point defects identified by the Wigner-Seitz defect analysis, with blue and red colors corresponding to interstitials and vacancies, respectively.

In these simulations, a constant PKA location was maintained across the different interfacial motifs for consistency. The rationale behind this decision is to minimize variability in simulation parameters when comparing the effects of interfacial structure on radiation damage. The PKA events are initiated below the interface for each case at an equal distance and averaging over multiple PKA atoms and configurations. It should be noted, however, that in a real radiation environment, cascades will be generated within the grain interior regions as well, producing defects that must diffuse through the bulk material in order to interact with interfacial sinks. MD simulations are not able to capture the longer time scales needed to effectively treat such behavior. The stress fields around these interfaces can also play a role in driving defect diffusion towards the interfaces, thereby influencing defect capture, diffusion barriers, and trapping strength. These additional complexities introduce a level of competition between defect generation and diffusion. Differences in simulation cell crystallography for the grain boundary and interphase boundary models mean that the displacement threshold energy may be slightly different, which can influence the defect production stage of the collision cascade. For this reason, only interfaces with the same crystallography, including amorphous complexions and their analogous ordered interfaces, are compared in absolute terms. However, since the PKA is initiated on the Cu side, given that all these atoms are Cu and possess the same mass, the variability in energies during the initial collisions should be minimal. It is acknowledged that inherent complexities of defect distribution within the collision cascade exist, where vacancies are often found concentrated in the core and interstitials are more prevalent around the periphery [62]. Due to this behavior, biases in the population of surviving defects might be introduced by the positioning of the cascade on each interfacial motif.

The interface models displayed in Fig. 6, while not intentionally, appear in order of their residual point defect content, with the single crystal model serving as the baseline state (i.e., zero sink strength) and the AIB model representing the configuration with the least amount of residual defect content. The single crystal model has an equal number of interstitials and vacancies, as no sink exists in the system and Frenkel pair recombination is the only available recovery mechanism. In contrast, the grain boundary model preferentially absorbs interstitials over vacancies, acting as a biased sink. For grain boundaries, recombination can come not only from vacancy diffusion but also from grain boundaries acting as interstitial sources upon irradiation [63]. The interphase boundary model also exhibits more residual vacancies than interstitials, although the difference is less significant than the ordered grain boundary. The mechanism behind irradiation damage recovery at the interphase boundary is associated with the free volume found at the near-coincident sites depicted in Fig 3, as previously shown by Demkowicz et al. [24] for similar interphase boundaries in other Cu-rich alloys. For example, the presence of more high excess volume sites in Cu-Nb compared to Cu-V correlated with an increased ability to resist He bubble formation [24,64]. The relaxation of point defects along these sites is responsible for the reduced residual defect content observed at the interphase boundaries, with misfit dislocation sites shown to be effective at trapping and delocalizing vacancies [65]. Notably, the damage is confined within the Cu grain where the PKA event originated, suggesting that the combination of interface structure and crystallographic orientation prevents damage from extending to the upper Ta grain. We remind the reader that the PKA was chosen to originate from the Cu side of the interface to allow for a comparison between the four different interface models. Initiating the PKA from the Ta grain would make any direct comparison impossible, yet we provide some limited results of such a simulation in Supplementary Note 1. Generally, we find that noticeable differences in the peak damage state exist depending on whether the PKA is initiated in the Cu or Ta crystals, yet the final damage states are similar.

Compared to the ordered interfacial structures, the increased interfacial thickness and structural disorder in the two types of amorphous complexions leads to less residual defect damage. Fig. 6 shows that the AGB and AIB models serve as more damage tolerant interfacial structures. However, it is important to clarify the limitations associated with MD simulations, such as the temporal constraints that restrict the possibility of all defects reaching the boundaries. This inherently means that overall sink efficiency is likely to be underestimated in such simulations, as long-range interactions are not fully captured. A major contributor to sink efficiency is defect diffusion to and absorption by the sink. Overlapping cascades with the interface only accounts for a limited set of scenarios, particularly situations where potential radiation damage is far away from the interface in question. When the AGB model is compared to its corresponding ordered counterpart (the grain boundary model), the inherent sink bias is maintained, as both are biased toward interstitials. This bias in the AGB model decreases as the thickness of the amorphous complexion increases, suggesting some influence of the confining Cu grains and their orientation in the sample with thinner complexions. However, the AGB model differs from the grain boundary model as it has a dramatically reduced defect content. Interestingly, the AGB in Cu-Ta serves as a slightly biased sink, whereas a comparable AGB in Cu-Zr was earlier noted to act as an unbiased sink [25]. The AIB model's extremely low residual defect content highlights its outstanding capability to absorb point defects. Overall, the damage healing behavior of the grain boundary, interphase boundary, and AGB is biased toward interstitials, whereas the AIB acts as an unbiased sink. In general, the amorphous complexions result in less residual damage compared to their analogous ordered interfaces examined in this study. Compared to the ordered grain boundary, the AGB shows a decrease in residual defect content of approximately 61% for interstitials and 64% for vacancies. In contrast, AIB exhibits reductions of ~83% for interstitials and ~93% for vacancies, respectively, as compared to the ordered interphase boundary.

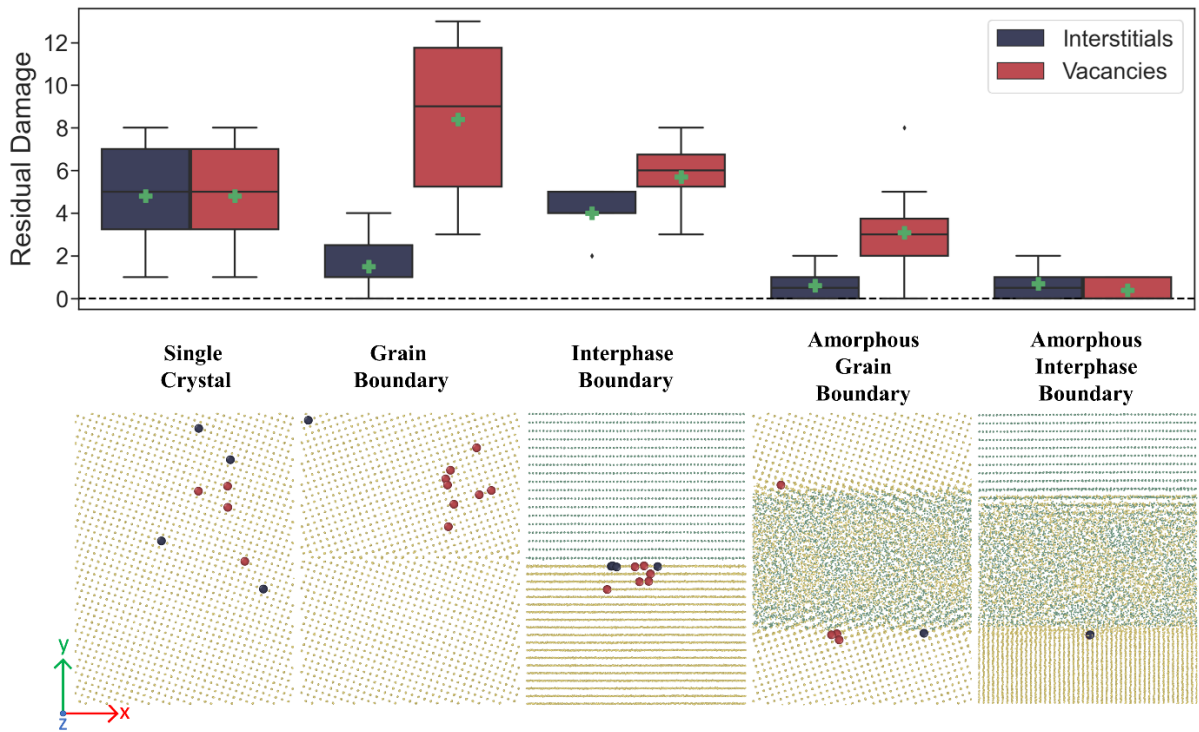

**Fig. 6.** Defect population metrics taken from 10 independent simulation runs for the single crystal, grain boundary, interstitial boundary, amorphous grain boundary (AGB), and amorphous interstitial boundary (AIB) models, with representative snapshots of the recovered configurations shown along the bottom row. Undamaged atoms are shown as small atoms, while larger particles correspond to either residual interstitial (colored blue) or vacancy (colored red) defects.

The average atomic volume of Cu atoms in the AGB and AIB regions are found to be 13.1%, and 11.7% higher compared to the non-defect Cu atoms in the models, respectively. The increased availability of free volume in amorphous complexions should contribute to defect sink efficiency, by altering important behaviors such as collision cascade confinement [25] and radiation-induced mixing [26]. The thickness of the amorphous complexions should also be important. As the thickness increased, amorphous grain boundary complexions in Cu-Zr were found to be more effective at absorbing dislocations during deformation [66], as they provide a larger region with free volume where plastic deformation could be shared. In addition, the disordered amorphous regions should provide more recombination opportunities, as opposed to structures that serve as pinning sites for interstitials in interfaces such as ordered grain boundaries [67]. Furthermore, diffuse interfaces have different defect production rate compared to sharper interfaces due to the different intrinsic displacement threshold energies [62].

To gain a deeper understanding of the effect of amorphous complexion thickness, the maximum and residual damage of AGB and AIB models with different film thicknesses was investigated. Fig. 7(a) shows snapshots from collision cascade simulations illustrating maximum and residual damage in the 2 nm thick AGB and AIB models. Figs. 7(b) and 7(c) show the number of defects from models with different film thicknesses for the maximum damage caused by the collision cascade and the residual damage, respectively. While the AGB and AIB results are illustrated in Fig. 7, differences in the maximum damage among the different interface types were also identified. It can be generally stated that SC and GB tend to exhibit higher maximum defect numbers, whereas IB shows lower defect generation during the collision cascade. The line plots and standard deviation error bars represent the number of interstitials (red) and vacancies (blue) of the 10 PKA simulations for the AGB and AIB models (solid and dashed lines). Focusing on Fig. 7(c), in general thicker amorphous complexions allow for the generation of fewer defects, although the effect starts to saturate for the thicker AGBs.

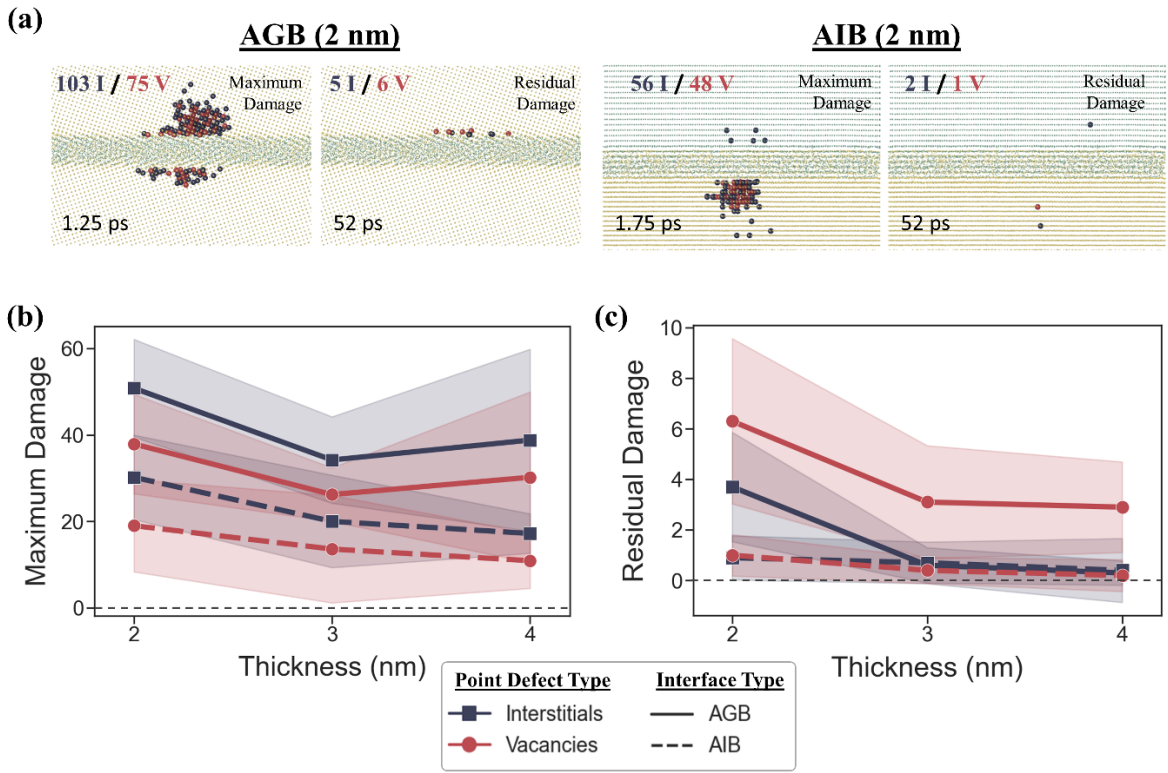

**Fig. 7. Comparative analysis of maximum and residual damage in AGB and AIB models of varying film thicknesses. (a) Collision cascade simulation snapshots depicting maximum and residual damage in 2 nm thick AGB and AIB models. The simulation time is indicated in the bottom left of each image, and the counts of interstitials (blue) and vacancies (red) are presented in the upper left. (b) The maximum damage resulting**

**from the collision cascade and (c) the residual damage at the end of the simulations for AGB and AIB models with different thicknesses.**

As amorphous complexion thickness increases, the point defects created by the PKA are more likely to be located inside a structurally disordered region, which should impact the radiation damage tolerance of an interface type by placing the defects closer to structurally heterogeneous sites. For instance, amorphous complexions will facilitate atom mobility, particularly due to the presence of relaxation pathways resulting from the fragmented polyhedral clusters seen in metallic glasses [68], as shown by prior studies of deformation in Cu-Zr glasses [69,70]. These regions, characterized by higher configurational entropy and lower symmetry, enable localized shear flow. This enhanced mobility provides an environment where radiation-induced defects can recombine more rapidly, which is amplified in thicker films owing to the longer diffusion lengths and higher concentration gradients, ultimately improving the material's radiation tolerance [71]. The nanophase separated structure of Cu-Ta amorphous complexions has important implications for radiation damage tolerance. Our simulations suggest that only minor structural adjustments occur post-collision cascade (see Supplementary Note 2). This lack of evolution can be attributed to the competition between defect generation and diffusion processes, which are accelerated due to the nanophase structure's inherent higher free volume as compared to a more homogeneous glass. Interestingly, we find that collision cascades introduce minimal structural alterations in both AGB and AIB regions across all thicknesses, which can be seen in Supplementary Figs. 2 and 3. For the quenched models shown in Fig. 4(a) and (b), the free volume in Cu-Ta was notably higher than in Cu-Zr amorphous complexions, substantiating the claim that nanophase separation and the resulting higher free volume play a role in its increased radiation damage tolerance.

Fig. 7(a) also demonstrates that there appears to be a difference in terms of where the damage is generated for the AGB and AIB samples, with damage occurring roughly equally in both grains for the AGB but with noticeably more damage in the bottom grain for the AIB. One could hypothesize that this behavior is also thickness dependent, with thicker films predicted to have a decrease in the damage propagation across the interface. To probe this concept, the percentage of damage that crossed interfaces into the upper grain during the collision cascade simulations is investigated. Fig. 8(a) shows the average residual damage distribution, broken into the crossing and non-crossing damage in terms of percentage. Fig. 8(b) illustrates four examples from the four different interface types, where a red line denotes the position that must be crossed for a defect to be considered as having crossed the interface. The choice of the upper amorphous-crystalline boundary for the amorphous complexion models is not critical, as this measurement could be taken with reference to the lower boundary and the exact same measurements would be extracted, as defects within the interface are not considered in the analysis (previously discussed in the Methods section). The residual damage in ordered grain boundaries predominantly crosses the interface, while the residual damage for the AGB and AIB models is primarily contained in the grain where the PKA originated. Generally, the amorphous complexions are better than their ordered counterparts at keeping damage in the bottom grain where the PKA originates, with thicker complexions indeed better at accomplishing this outcome.

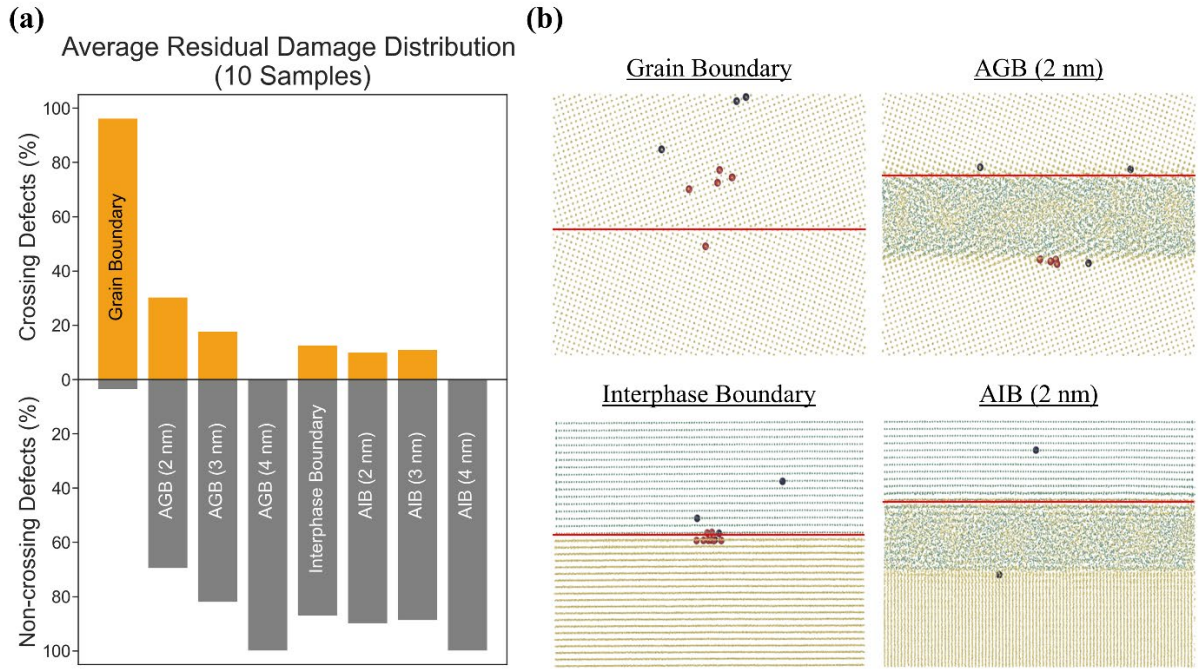

Fig. 8. (a) The spatial distribution of average residual damage, separated into interface crossing and non-crossing defect populations. (b) Snapshots from four representative simulations of the four interface types, where the red line within the snapshots delineates the boundary that a defect must cross to be categorized as being "interface crossing."

Finally, to better understand the limits of the observed damage tolerance for the amorphous complexions, the effect of increasing PKA energy is investigated for the 3 nm thick AIB (selected due to its place as the most damage tolerant interface type). Higher doses of radiation result in more ions with kinetic energies exceeding the threshold displacement energy [41], which also corresponds to the more extreme environments expected in future advanced fission and fusion reactors, as well as the ion irradiation energies used to emulate these environments. Fig. 9 shows the maximum damage and residual damage configurations in the 3 nm thick AIB model for three different PKA energies. The 5 keV PKA energy generates much larger maximum damage than the original 2 keV PKA, yet the damage healing efficiency is maintained and the final configurations are similar with very few residual point defects. Further increasing the PKA energy to 10 keV results in much higher point defect content, reaching over a thousand interstitials and vacancies, which is high enough that the point defect field can relax into higher order defect types. A collection of dislocation loops is formed, with seven Shockley partial dislocations with $\vec{b} = {a}/{6}\langle 112 \rangle$ identified by the DXA tool in OVITO [53]. Fig. 9(c) shows this residual damage configuration, which is still primarily contained within the Cu phase, consistent with the earlier observations at lower energies. As discussed previously, excess free volume acts as a catalyst for enhancing recombination. Despite the complex defect landscape and the significant number of defects generated at higher PKA energies, the AIB displays the ability to reconfigure and absorb defects, which likely involves the annihilation of defects at the interface or their clustering. The surplus of 18 interstitials in Fig. 9(c) suggests that the interface was unable to relax those residual defects when subjected to a high PKA energy, serving as a signal that its ability to do so is being overwhelmed. The unique nanophase separation observed in Cu-Ta systems further complicates this balance, as it likely represents a metastable state that could affect both the availability and utility of free volume. After several recombinations facilitated by the AIB, it is possible that the nature of the excess free volume undergoes alterations, resulting in an inadequate mean free path associated with defect absorption to halt the collision cascade. This effect may be even more pronounced for the AGB, given its interstitial bias. The agglomeration and interaction of defects with each other that leads to the formation of defect clusters or even dislocations are a common experimental observation. At homologous temperatures greater than 30% of the absolute melting temperature, radiation-induced defects are generally mobile [72]. For instance, a study investigating defect migration behavior in Ni, Ni-Fe, and Ni-Co using ion irradiation observed the migration of interstitial clusters, a collection of 1-D clusters primarily migrating along the <110>

direction, into deeper regions and eventually the creation of a network of dislocations [3]. As a whole, this emphasizes that while amorphous complexions exhibit reduced residual damage content, they can still be overwhelmed if too many point defects are added too quickly to the system.

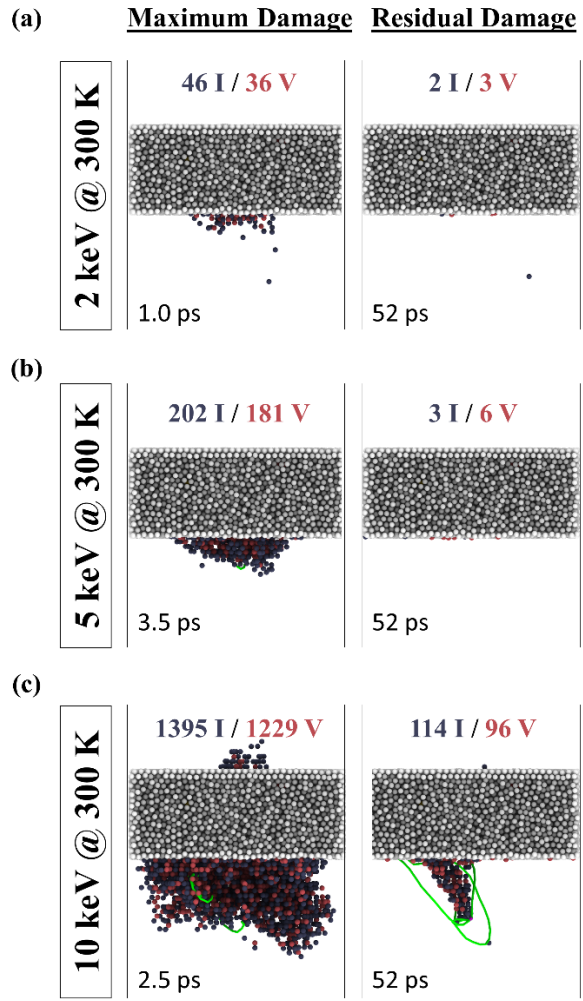

**Fig. 9. Atomic snapshots of the 3 nm thick AIB model showing the maximum (left) and residual damage (right) states from collision cascade simulations with PKA energies of (a) 2 keV, (b) 5 keV, and (c) 10 keV. Blue and red colored atoms represent interstitials (I) and vacancies (V), respectively, while the green colored lines denote dislocations and perfect crystalline atoms are not shown.**

## 4. CONCLUSIONS

In this work, the radiation damage resistance of amorphous complexions in the Cu-Ta alloy system was investigated using MD simulations. Collision cascades from PKA events were simulated for different interface types, including a grain boundary, an interphase boundary, an amorphous grain boundary complexion, and an amorphous interphase boundary complexion. The effect of interfacial disorder on point defect healing and residual defect content was studied, as well as the importance of complexion thickness on damage resistance. Key findings from this work include:

(1) The nanophase separated structure of amorphous complexions in Cu-Ta affects the healing of radiation damage, as compared to alloys with higher glass-forming ability such as Cu-Zr. For example, the AGB in Cu-Ta acts as a biased sink, while a similar AGB in Cu-Zr was previously observed to be an unbiased sink.

(2) The amorphous complexion models, AGB and AIB, exhibit reduced residual defect content compared to their analogous ordered boundaries. The radiation damage healing behavior of the grain boundary, interphase boundary and AGB were biased toward interstitials, while the AIB models acted as an unbiased sink with outstanding efficiency.

(3) Free volume availability is found to contribute to the minimal residual damage exhibited by the amorphous complexions. This behavior is attributed to the disordered amorphous regions contributing to collision cascade confinement and promoting defect recombination and annihilation due to higher excess atomic volume.

(4) The thickness of amorphous complexions noticeably impacts their damage tolerance. Thicker films generate fewer defects and exhibit lower residual point defect content.

Increased thickness exposes mobile defects to disordered regions for longer, enhancing defect migration and recombination and reducing damage propagation across the interface.

(5) Even the most damage tolerant interfaces can be overwhelmed if the PKA energy is too high (i.e., the point defect content is too large). The AIB stopped being able to completely relax damage when high-energy collision events were simulated, which can be primarily attributed to the exhaustion of recombination sites.

In summary, amorphous interfacial complexions within Cu-Ta show minimal residual defect damage compared to ordered interfaces such as grain boundaries and interphase boundaries. These results suggest that radiation damage-resistant materials can be engineered through microstructural design by increasing the fraction and thickness of amorphous interfaces in Cu-Ta alloys. Thermomechanical treatments [73,74] and planned grain boundary doping [75–77] have been used to achieve these types of results in other alloys. However, the ability to tailor interfacial structure in a controlled manner through specific irradiation conditions remains a future goal, with key factors such as the onset dose for amorphization and the effect of temperature being currently unknown. The current study creates the foundational for such future work, by demonstrating the beneficial properties of amorphous complexions in Cu-Ta. Since these features in fact form during irradiation, they represent a possible pathway for the in operando creation of damage tolerant microstructural features under service conditions relevant to nuclear reactors and other high-radiation environments.

**DATA AVAILABILITY STATEMENT**

The data that supports the findings of this study are available within the article.

**ACKNOWLEDGEMENTS**

This research was primarily supported (atomistic simulations by D.A.) by the National Science Foundation Materials Research Science and Engineering Center program through the UC Irvine Center for Complex and Active Materials (DMR-2011967). Work by T.J.R. was supported by the U.S. Department of Energy, Office of Science, Basic Energy Sciences, under Award No. DE-SC0021224 (theory and analysis of grain boundary complexions). Work by J.P.W. was supported by the U.S. Department of Energy, Office of Science, Basic Energy Sciences, under Award No. DE-SC0020150 (irradiation tailoring of phase transformations). Experimental results from ion irradiated Cu-Ta were obtained at the Intermediate Voltage Electron Microscope (IVEM)-Tandem Facility at Argonne National Laboratory through the U.S. Department of Energy, Office of Nuclear Energy, Nuclear Science User Facilities, under Award No. 19-1757. J.R.T. acknowledges support from the U.S. Department of Energy, Office of Science, Basic Energy Sciences, under Award No. DE-SC0021060. P.C. was supported by the U.S. Department of Energy, Basic Energy Sciences, under Award No. DE-SC0022295.

**CONFLICT OF INTEREST**

All the authors declare that they have no conflicts of interest.

# Supplementary Materials

**Supplementary Note 1:**

The collision cascade events shown in the main body of the text were consistently initiated from the Cu grain side. Here, an alternative initiation of the collision cascade event from the Ta grain was explored. It was observed that while the peak damage states displayed significant differences depending on the PKA initiation point, the final damage states converged to similar configurations, which are shown in Fig. S1. Notably, when the PKA was initiated from the Ta side, a larger peak damage was observed, although the final damage state remained similar to that observed after PKA initiation in the Cu grain. The findings presented in Fig. 9 of the main text indicate that damage confinement within the Cu grain is influenced by both the PKA energy and its proximity to the interface, with higher PKA energies potentially leading to saturation at the interface. Considering the interplay of these factors affecting the evolution of the collision cascade, a consistent approach of initiating the PKA in the Cu grain was adopted across all interfaces to facilitate a balanced and comparable analysis.

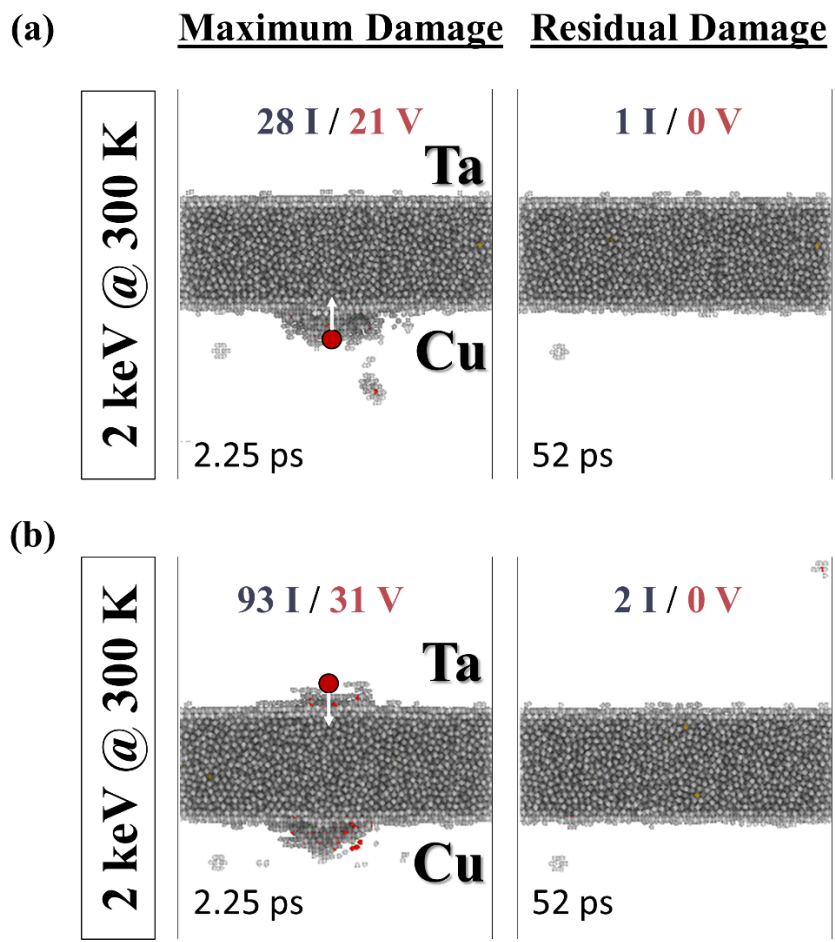

**Fig. S10.** Atomic snapshots of the AIB model showing the maximum (left) and residual damage (right) states from collision cascade simulations with a PKA energy of 2 keV. These snapshots compare scenarios where the collision cascade is initiated (a) from the Cu side and (b) from the Ta side, providing insight into the differing damage profiles. The PKA atom is highlighted in red, with the white arrow indicating its initial direction of movement. Perfect crystalline atoms are omitted for clarity.

**Supplementary Note 2:**

An investigation into the local structural evolution of the amorphous complexions reveals that these features exhibit only minor structural adjustments following collision cascade events. The resilience of these complexions to radiation-induced changes is depicted by radial distribution functions for the AGB and AIB models in Figs. S2 and S3, respectively for varying complexion thicknesses, before and after the collision cascade. No significant changes are observed in the radial distribution functions due to radiation damage. The observed stability corroborates the hypothesis that the nanophase structure's higher free volume facilitates a balance between defect generation and diffusion processes, thereby enhancing the material's radiation tolerance.

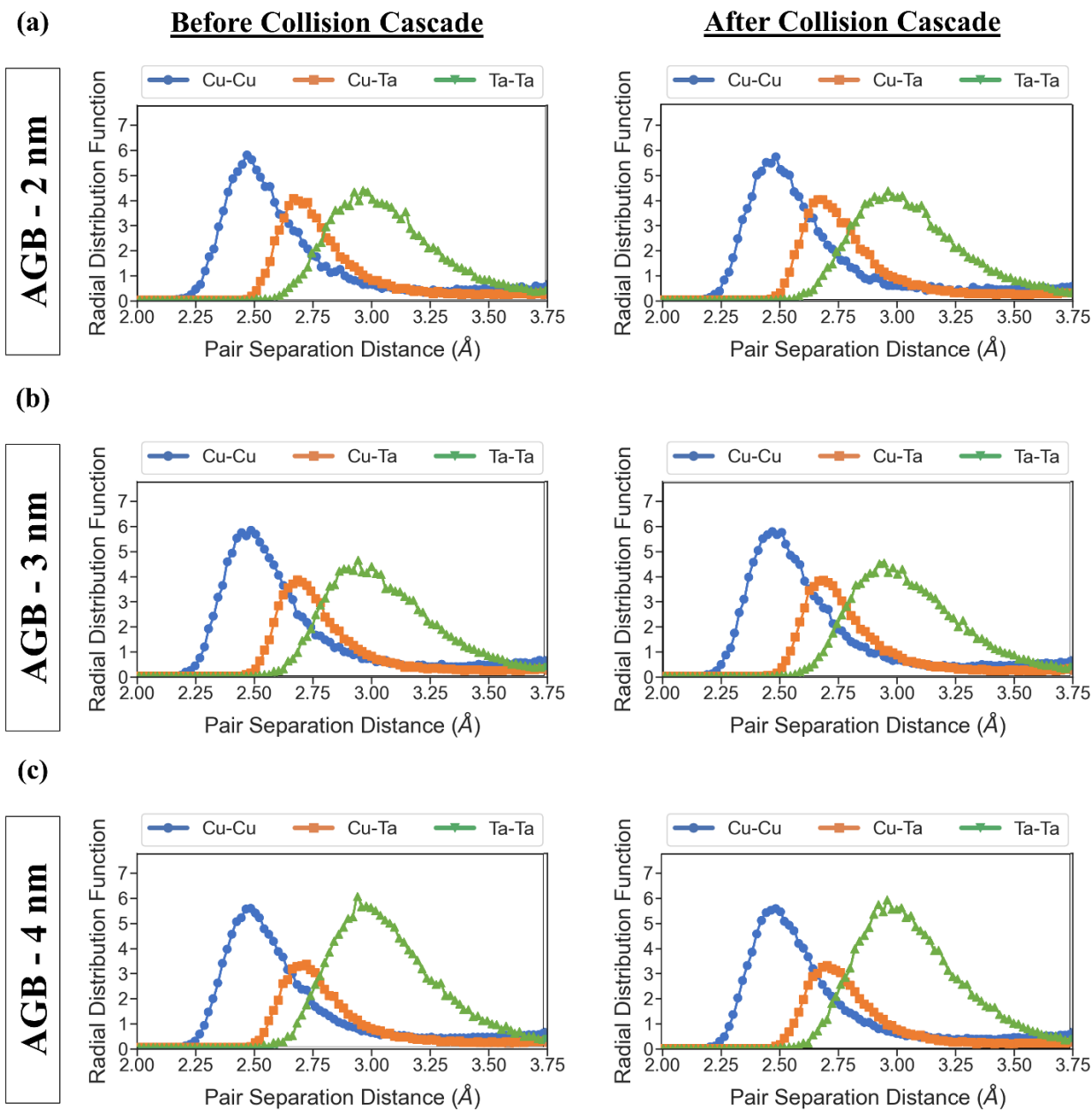

**Fig. S11.** Radial distribution functions of Cu-Cu, Cu-Ta, and Ta-Ta pairs in the AGB model as a function of pair separation distance, measured for models with (a) 2 nm, (b) 3 nm, and (c) 4 nm complexion thicknesses, before (left column) and after (right column) collision cascade events. The comparison of the pre- and post-damage scenarios across various thicknesses illustrates the minimal structural disturbances caused by radiation.

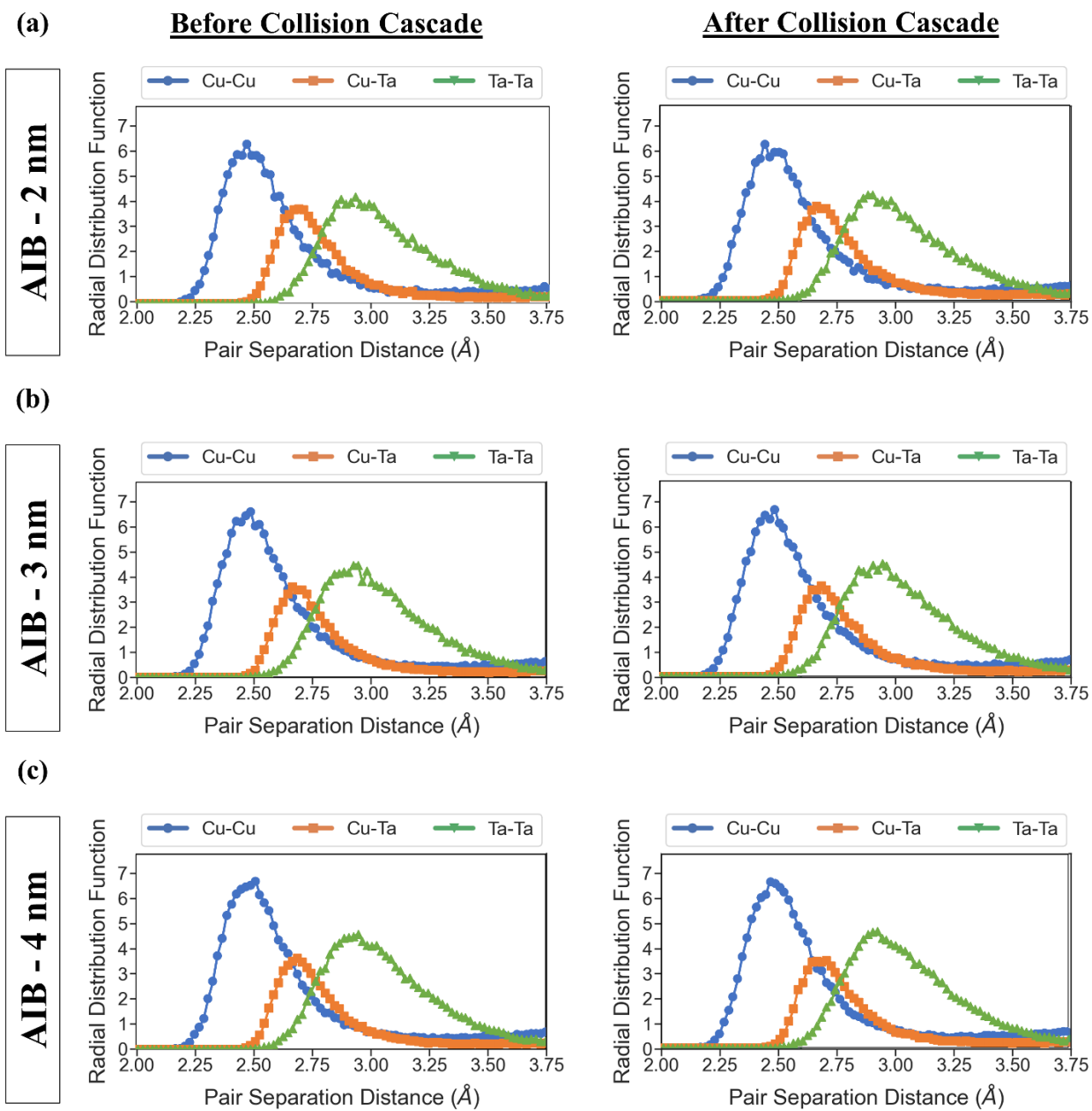

**Fig. S12. Radial distribution functions of Cu-Cu, Cu-Ta, and Ta-Ta pairs in the AIB model as a function of pair separation distance, measured for models with (a) 2 nm, (b) 3 nm, and (c) 4 nm complexion thicknesses, before (left column) and after (right column) collision cascade events. The comparison of the pre- and post-damage scenarios across various thicknesses illustrates the minimal structural disturbances caused by radiation.**